\pdfoutput=1
\documentclass[aps,prl,reprint,superscriptaddress]{revtex4-2}

\usepackage{amsmath}
\usepackage{svg}
\usepackage{siunitx}
\definecolor{bluepoli}{RGB}{0,36,179}
\definecolor{redpoli}{RGB}{204,0,51}
\definecolor{greenpoli}{RGB}{45,137,0}
\definecolor{purplepoli}{RGB}{153,102,204}
\definecolor{azzurropoli}{RGB}{51,53,204}
\definecolor{orangepoli}{RGB}{255,124,17}
\usepackage{url}
\usepackage{hyperref}
\usepackage{physics}
\usepackage{upgreek}
  
\hypersetup{
	colorlinks,
	linkcolor={bluepoli}, %blue poli
	citecolor={redpoli}, %rosso
	urlcolor={redpoli} % 
}

\begin{document}

\title{A quantum dot in germanium proximitized by a superconductor}

\author{Lazar Lakic$^\dagger$}
\affiliation{Center for Quantum Devices, Niels Bohr Institute, University of Copenhagen, 2100 Copenhagen, Denmark}

\author{William Iain L.~Lawrie$^\dagger$}
\affiliation{Center for Quantum Devices, Niels Bohr Institute, University of Copenhagen, 2100 Copenhagen, Denmark}

\author{David van Driel}
\affiliation{
QuTech and Kavli Institute of Nanoscience, Delft University of Technology, Delft, The Netherlands}

\author{Lucas E.~A.~Stehouwer}
\affiliation{
QuTech and Kavli Institute of Nanoscience, Delft University of Technology, Delft, The Netherlands}

\author{Yao Su}
\affiliation{Center for Quantum Devices, Niels Bohr Institute, University of Copenhagen, 2100 Copenhagen, Denmark}

\author{Menno Veldhorst}
\affiliation{
QuTech and Kavli Institute of Nanoscience, Delft University of Technology, Delft, The Netherlands}

\author{Giordano Scappucci}
\affiliation{
QuTech and Kavli Institute of Nanoscience, Delft University of Technology, Delft, The Netherlands}

\author{Ferdinand Kuemmeth}
\affiliation{Center for Quantum Devices, Niels Bohr Institute, University of Copenhagen, 2100 Copenhagen, Denmark}

\author{Anasua Chatterjee}
\email{anasua.chatterjee@tudelft.nl}
\affiliation{Center for Quantum Devices, Niels Bohr Institute, University of Copenhagen, 2100 Copenhagen, Denmark}
\affiliation{
QuTech and Kavli Institute of Nanoscience, Delft University of Technology, Delft, The Netherlands}

\date{\today
}
	
%:abstract	
\begin{abstract}

Planar germanium quantum wells have recently been shown to host {hard-gapped superconductivity}. Additionally, quantum dot spin qubits in germanium are well-suited for quantum information processing, with isotopic purification to a nuclear spin-free material expected to yield long coherence times. {Therefore, as one of the few group IV materials with the potential to host superconductor-semiconductor hybrid devices, proximitized quantum dots in germanium is a compelling platform to achieve and combine topological superconductivity with existing and novel qubit modalities.} Here we demonstrate a quantum dot (QD) in a Ge/SiGe heterostructure proximitized by a platinum germanosilicide (PtGeSi) superconducting lead (SC), forming a SC-QD-SC junction. We show tunability of the QD-SC coupling strength, as well as gate control of the ratio of charging energy and the induced gap. We further exploit this tunability by exhibiting control of the ground state of the system between even and odd parity. Furthermore, we characterize the critical magnetic field strengths, finding a critical out-of-plane field of {$0.90~\pm0.04$~T}. Finally we explore sub-gap spin splitting in the device, observing rich physics in the resulting spectra, that we model using a zero-bandwidth model in the Yu-Shiba-Rusinov limit. The demonstration of controllable proximitization at the nanoscale of a germanium quantum dot opens up the physics of novel spin and superconducting qubits, and Josephson junction arrays in a group IV material.
\end{abstract}

\maketitle

%:Introduction	
\section{Introduction}

New and exotic physical phenomena can emerge in superconducting-semiconducting hybrids, enabling engineered quantum materials~\cite{Boettcher2018}, circuit quantum electrodynamics (cQED) with novel superconducting qubits~\cite{Hays2021,PitaVidal2023}, and topologically protected phases~\cite{LeijnseParity,Kitaev,tenHaaf2023,Dvir2023}. In particular, proximitized quantum dots constitute key building blocks for devices such as Cooper pair splitters~\cite{Wang2022}, Kitaev chains~\cite{tenHaaf2023,Dvir2023}, and protected qubits~\cite{Pino2024,Wang2023}. However, to date the majority of these experiments have been performed in group III-V materials where nuclear spins are unavoidable, critically hampering spin coherence, and where 2D heterostructures exhibit piezoelectricity, deleterious for cQED circuits~\cite{Scigliuzzo2020}. Conversely, silicon and germanium are established group IV material platforms to integrate spin qubits hosted in gate defined quantum dots~\cite{BurkardSpinQubits,Scappucci2021}, with isotopic purification having proved {crucial} for ultra-long spin qubit coherence~\cite{Muhonen2014}. Contrary to silicon~\cite{MengTao2003}, germanium forms low resistance Ohmic contacts due to intrinsic Fermi level pinning close to the valence band~\cite{Dimoulas2006,Nishimura2007}. This has motivated a strong effort to induce superconductivity~\cite{Aggarwal2021, Vigneau2019, Hendrickx2018, Ridderbos2018, Tosato2023, Valentini2024} and very recently, {signs of hard-gap superconductivity have been observed in mesoscopic devices implemented in a Ge/SiGe heterostructure~\cite{Tosato2023,Valentini2024}, in Ge/Si core shell nanowires~\cite{Ridderbos2019,Zhuo23,Zheng2024} and in a cQED circuit~\cite{Hinderling2024}.} 

Here, we present a superconducting-semiconducting hybrid quantum dot, which is hosted in Ge, a group IV material uniquely allowing for both isotopic purification~\cite{Itoh1993} and a superconducting hard gap~\cite{Tosato2023,Valentini2024}. Our demonstration in a two-dimensional heterostructure establishes a novel platform that exhibits enhanced scalability compared to nanowires, is compatible with radiofrequency-reflectometry readout~\cite{Vigneau2022} and a highly successful spin qubit platform~\cite{Borsoi2024,Hendrickx2021,Jirovec2021}. It may therefore be useful for {extending the range of qubit interactions, by using} crossed Andreev reflection~\cite{LeijnseUndKarsten,Spethmann2024} as well as heterogeneous quantum processors~\cite{Pino2024,Gyenis2021} incorporating spin ~\cite{BurkardSpinQubits,Scappucci2021} and superconducting circuits~\cite{scqubits}. Isotopically purified, proximitized germanium {may enable} coherent Andreev spin qubits, protected superconducting qubits, and quantum dot-based Kitaev chains, and our demonstration of a quantum dot with gate-tunable proximitization in a group IV heterostructure is a key ingredient.

\begin{figure*}
		\includegraphics{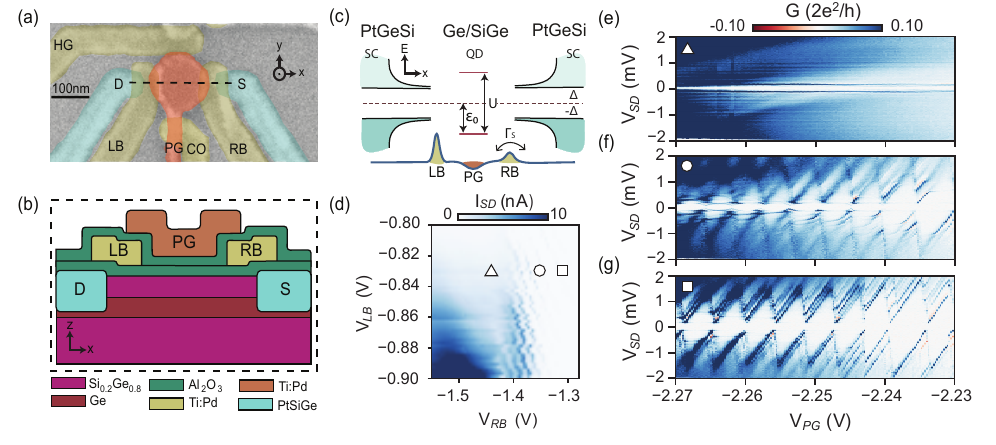}
		\caption{\textbf{(a)} False coloured scanning electron micrograph (SEM) of a nominally identical device. The device comprises three lithographically defined metallic layers, separated by a dielectric of Al$_2$O$_3$ grown using atomic layer deposition (ALD). A plunger gate~PG (orange) controls the electrochemical potential of the quantum dot. The coupling of the quantum dot to the superconducting PtSiGe leads S and D is controlled by two barrier gates LB and RB (yellow). A cut-off gate CO prevents accumulation beneath the gate fan-out of PG, and a helper gate HG provides further control of the quantum dot confinement. \textbf{(b)} Heterostructure and gate stack schematic corresponding to the cross section indicated by the black dashed line in (a). \textbf{(c)} Energy schematic depicting the physical system in (a). Here, $\Delta$ is the SC gap energy, $U$ the charging energy of the QD, $\Gamma_\text{S}$ the hybridization energy of the SC and QD and $\epsilon_0$ the electrochemical potential of the QD with respect to the SC Fermi energy. \textbf{(d)} Source-drain current I$_\text{SD}$ as a function of barrier gates $V_\text{LB}$ and $V_\text{RB}$ at bias voltage $V_\text{SD}$ = {300~$\upmu$V}, {$V_{PG}=-2.22$~V, and $V_{HG}=-1.00~V$}. The square, circle and triangle correspond to the indicated gate voltage setting in (e), (f) and (g). {Horizontal features likely correspond to charge instabilities in the environment surrounding the quantum dot.} \textbf{(e-g)} Bias spectroscopy for the three gate voltages indicated in (d), with high, moderate and low coupling of the quantum dot to the superconducting leads respectively, showing a transition between strongly coupled lead (e) and weakly coupled lead (g). Negative differential conductance observed may indicate Couloumb diamonds of odd occupancy \cite{DeFranceschi2010}. 
	}
        \label{fig:Figure.1}
\end{figure*}

In this work, superconducting polycrystalline Platinum-Germanium-silicide (PtSiGe) leads are formed by a controlled thermally-activated solid phase reaction between deposited platinum (Pt) and the heterostructure (Ge/SiGe)~\cite{Tosato2023}. Importantly, the PtGeSi leads alleviate the need to etch into the heterostructure to deposit or pattern the superconductor, a potential source of damage exposing the quantum well and interface to {oxygen} and processing. {This method of achieving a transparent interface to a superconductor, via a reaction with a deposited noble metal, constitutes a general technique that has been exploited in other materials~\cite{Bai2020,Rosen2024,Jia2024}}. The leads act as charge reservoirs and {as a source of} proximitization for the quantum dot (QD). We first demonstrate Coulomb blockade physics of a QD coupled to two superconducting leads (SC) forming a SC-QD-SC junction. We identify a superconducting gap energy window of $4\Delta_0$ {= 284 $\mu$eV} inside which transport is suppressed, and outside which standard Coulomb diamonds are recovered. {The charging energy of the system is typically larger than 1 meV, making it the dominating energy scale. Consequentially, Yu-Shiba-Rusinov~(YSR) physics is expected to describe the observed phenomena}. We observe sub-gap states in transport upon increasing the QD-SC coupling $\Gamma_\text{S}$, which is consistent with the formation of {YSR} states in the system~\cite{Kirsanskas2015, PhysRevB.94.064520}. We demonstrate gate control of $\Gamma_\text{S}$ by tuning the ground state at half-filling from a singlet state to a doublet state~\cite{PRXQuantum}. We then study the critical magnetic field of the hybrid device, finding an out-of-plane critical magnetic field {$B_{\perp}^c = 0.90~\pm~0.04$~T}. Finally, we investigate spin-splitting in the SC-QD-SC system, finding a $g$-factor of 1.5$~\pm~0.2$ for an out-of-plane magnetic field, and also characterize the $g$-tensor anisotropy. To explain the energy splitting observed in the SC-QD-SC system we use a zero bandwidth (ZBW) Anderson Impurity model~\cite{Bauer_2007,meng2009self} with the possibility of Zeeman splitting on the SC. Our observation of controllable subgap states and subgap spin splitting, magnetic field resilience, and the high tunability of the quantum dot-superconducting coupling establishes Ge/SiGe and PtSiGe as an attractive platform for hybrid quantum information processing.

\begin{figure*}
    \includegraphics{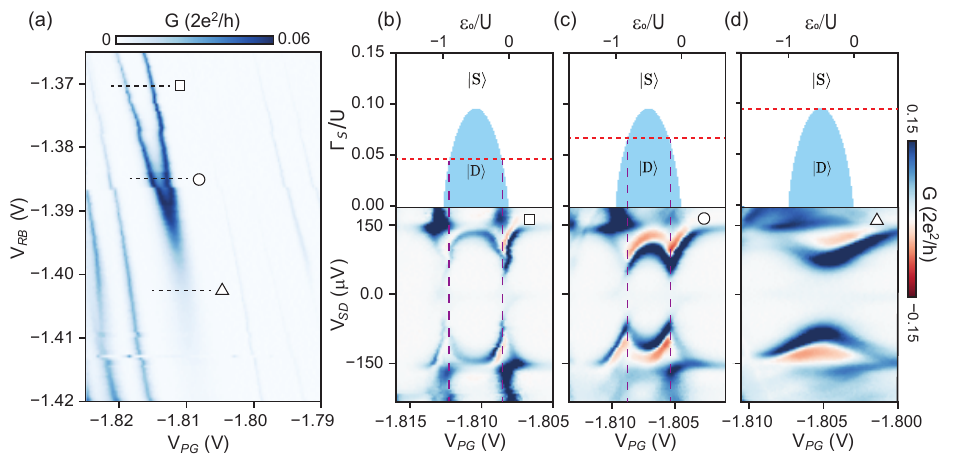}
    \caption{\textbf{(a)} Charge stability diagram of $V_\mathrm{RB}$ vs $V_\mathrm{PG}$ at $V_\mathrm{SD}=80$~$\upmu$V{, $V_\mathrm{LB}$=-1.108~V, and $V_{HG}=-1.000$~V}. {A cross-capacitance between the elecrostatic gate V$_{RB}$ and the quantum dot results in simultaneous tuning of $\Gamma_S$ and $\epsilon_0$. We note that we expect the quantum dot system to be in the multi-hole regime.} \textbf{(b-d)} Bottom panels show bias spectroscopy at decreasing values of $V_\mathrm{RB}$ corresponding to the square, circle, and triangle icons in panel a. {Upper panels portray phase diagrams for the singlet and doublet phases, computed using a minimal ZBW model of the expected ground state character of the hybrid system. Here, $\epsilon_0$ is the electrochemical potential of the quantum dot with respect to the grounded superconducting leads, and $U$ is the charging energy of the quantum dot, the phase diagram was computed assuming $U$=1.6~meV and {$\Delta=71~\upmu$eV}}. As the SC-QD coupling $\Gamma_\mathrm{S}$ is increased, the doublet state becomes energetically unfavorable, as seen by the merging of charge transitions (magenta dashed lines). $\Gamma_\mathrm{S}$ is roughly estimated by modelling the sub-gap spectrum in the bottom panels (see Supplementary Information {section V}), to be 70~$\upmu$eV, 110$~\upmu$eV, and 150~$\upmu$eV from b-d respectively, as indicated by the red dashed lines.
    } 
    \label{fig:Figure.2}
\end{figure*}

\section{Results}	
%:Figure 1	
We utilize established fabrication protocols for {QD} fabrication~\cite{Lawrie2020} and superconducting contacts~\cite{Tosato2023} in Ge/SiGe quantum wells~\cite{Sammak2019}, to create a quantum dot coupled to two superconducting leads formed by rapid thermal annealing of Pt at $400^{\circ}$C in Ar atmosphere for 15 minutes. Figure \ref{fig:Figure.1}a  shows a false-colored scanning electron micrograph of the device, consisting of one lithographically-defined layer for the superconducting leads (cyan) and two layers of electrostatic gates (yellow and orange, see Methods and section I for further details). Figure \ref{fig:Figure.1}b shows a schematic of the cross-section of the device heterostructure and gate stack. Layers are electrically isolated from one another by 7\,nm of Al$_2$O$_3$ deposited by atomic layer deposition at $150^{\circ}$C. Barrier gates (LB,~RB) control $\Gamma_\text{S}$, while the plunger gate (PG) controls the relative electrochemical potential of the quantum dot levels with respect to the superconducting leads ($\epsilon_0$) as seen in Figure \ref{fig:Figure.1}c. Two gates (HG and CO) are also used to confine~{(HG)} the quantum dot and prevent unwanted accumulation~{(CO)}. We utilize standard DC transport and low frequency lock-in techniques to measure source-drain current $I_\mathrm{SD}$ and differential conductance $G$ across the quantum dot. Notably, our device is also connected to a radiofrequency (RF) reflectometry circuit via the source superconducting lead, {both techniques described in Supplementary Information section I}. Additional datasets in the low-coupling regime measured using this RF probe are presented in the Supplementary Information {section II}.\\
\newline
All data are taken at a lock-in frequency of 119~Hz, and amplitude of 2.5~$\upmu$V. Figure \ref{fig:Figure.1}d shows $I_\mathrm{SD}$ as a function of the tunnel barrier $V_\mathrm{LB}$ and $V_\mathrm{RB}$. Here, the source drain bias energy is set to $eV_\mathrm{SD}$ = 300 $\upmu$eV such that it exceeds the expected zero-field superconducting gap energy of $\sim70\,\,\upmu \mathrm{eV}$~\cite{Tosato2023}. We set $V_\mathrm{LB}$ close to its pinch-off value such that it acts as a tunnel probe, and vary $V_\mathrm{RB}$ to tune the coupling between superconductor and quantum dot $\Gamma_\mathrm{S}$ (Figure \ref{fig:Figure.1}c). {This limits current through the device to around 1 nA, which we estimate to be well within the critical current density of the superconducting leads (See Supplementary Figure S6b).} Figures 1e-g show bias spectroscopy at different values of $V_\mathrm{RB}$. In the strong coupling regime (Figure \ref{fig:Figure.1}e), we observe a range in bias energy of $4\Delta_0$ where transport is suppressed, from which we extract a superconducting pairing amplitude of $\Delta_0$ = {71~$\pm$~6}~$\upmu$eV, in its fully open state. {The apparent reduction of the SC gap upon increased $V_{PG}$ is attributed to sub-gap transport phenomena due to increased junction transparency~\cite{Morten}}. {We furthermore note that the dark features outside of the SC gap likely result from out of gap structure in the SC leads~\cite{Mirage}}. Figure \ref{fig:Figure.1}f shows that as $\Gamma_\mathrm{S}$ is decreased (positive change on $V_\mathrm{RB}$), we observe tunnel-broadened Coulomb oscillations and sub-gap transport features, indicating a hybridized QD. 
At low coupling between the QD and SC (Fig. \ref{fig:Figure.1}g) we observe sharp Coulomb diamonds outside a bias window of $\pm$2$\Delta_0$. 
We conclude that we have versatile electrostatic control of the degree of hybridization of a quantum dot with a superconductor, consistent with experiments performed in InAs-Al nanowires~\cite{Lee2014} and InSbAs-Al 2DEGs \cite{tenHaaf2023}.

\subsection{Singlet-doublet quantum phase transition}
A quantum dot coupled to a superconducting lead at half-occupancy of charges can have two different ground states, depending on the degree of superconductor-quantum dot coupling. At low coupling strengths and zero magnetic field strength, the ground state at half-filling,  ie. $\epsilon_0/U = 0.5$, will be a spin-degenerate doublet state $\ket{D}=\{ \ket{\downarrow}, \ket{\uparrow} \}$. Here, $\epsilon_0$ is the electrochemical potential of the QD with respect to the SC lead and $U$ is the charging energy of the QD. At high coupling, a preference for superconducting pairing will dominate, leading to a singlet ground state $\ket{S}=u\ket{0}-v\ket{2}$. {We stress that the labeling of these states is not representative of the localization of the charges between superconductor and quantum dot, and rather a convenient pedagogical representation of the basis states of the system. As a result of $U$ being the dominant energy scale, the system is firmly in the YSR limit.} By utilizing the control of $\Gamma_\text{S}$ demonstrated above, we show that we can tune between these ground states.
We operate in a regime whereby $V_\mathrm{LB}$ is very close to its pinch-off value, such that it acts as a tunneling probe. We then vary $V_\mathrm{RB}$ to tune the coupling $\Gamma_\mathrm{S}$. Figure \ref{fig:Figure.2}a shows a charge stability diagram of the system with the QD plunger gate $V_\mathrm{PG}$ on the horizontal axis, and the QD-SC barrier gate $V_\mathrm{RB}$ on the vertical axis, at a bias energy of  $eV_\mathrm{SD} =80$~$\upmu$eV, slightly above $\Delta$.
The vertical lines measured are Coulomb resonances indicating transitions between the $N$, $N+1$ and $N+2$ occupations of the quantum dot (from right to left). As we increase $\Gamma_\mathrm{S}$ by making RB more negative we observe the merging of two levels at {approximately} $V_\mathrm{RB}$ = -1.395~V.

We further investigate these transitions with bias spectroscopy as a function of plunger gate voltage at different values of $V_\mathrm{RB}$. In the bottom panel of Figure~\ref{fig:Figure.2}b, we show bias spectroscopy at {$V_\mathrm{RB} = \SI{-1.3717}{\volt}$}~(square in Figure 2a) for varying $V_\mathrm{SD}$ and $V_\mathrm{PG}$.  At $V_\mathrm{PG}$ values of {-1.812~V} and -1.808~V, the state crosses $\Delta_0$, signalling the changes in ground state parity as seen from \ref{fig:Figure.2}a. 

In the bottom panel of Figure~\ref{fig:Figure.2}c, we show spectroscopy at $V_\mathrm{RB} = \SI{-1.3850}{\volt}$~(circle in Figure~\ref{fig:Figure.2}a). We see an evolution of the state features into a characteristic eye-shape indicating the formation of YSR states~\cite{Kirsanskas2015,Grove-Rasmussen2009,Yu65,Shiba,Rusinov68} on the hybridized QD. The negative differential conductance is attributed to probing sub-gap features with a coherence peak~\cite{NDC_Brian}.
In the bottom panel of Figure \ref{fig:Figure.2}d, we perform bias spectroscopy at {$V_\mathrm{RB} = \SI{-1.4025}{\volt}$}~(triangle in Figure~\ref{fig:Figure.2}a) showing no parity change, which we interpret as the QD filling remaining in a singlet ground state upon loading an additional hole.

In QD-SC systems, where the charging energy of the QD $U$ is larger than the SC order parameter $\Delta$, quasiparticles in the SC can bind to the dot by the exchange interaction and give rise to sub-gap excitations in the form of YSR states. Such systems can be modelled using a zero-bandwidth~(ZBW) model that describes a quantum dot coupled to a single superconducting orbital~\cite{Bauer_2007,baran2023surrogate}, which predicts which ground state the system prefers depending on the degree of hybridization between the SC and the QD. 
By solving the ZBW {model} for where the energy of the singlet equals the one of the doublet, a singlet doublet phase transition diagram can be realized. In the top panels of Figure \ref{fig:Figure.2}b-d such a phase transition has been {computed using a ZBW model with a charging energy of $U$ = 1.6~m$\mathrm{eV}$ and a superconducting gap of {$\Delta=71 ~\upmu\mathrm{eV}$}, the} dashed lines have been inserted at $\Gamma_\mathrm{S}$ values based on extracted coupling values ($70~\upmu\mathrm{eV}$, $110~\upmu \mathrm{eV}$, and $150~\upmu\mathrm{eV}$ from b-d respectively) {divided by \textit{U}}, {also} using the aforementioned minimal ZBW model~(see Supplementary Information {section V}). As the hybridization energy $\Gamma_\mathrm{S}$ increases, it becomes less favorable to maintain the $\ket{D}$ ground state. The {horizontal} dashed lines thus indicate a qualitative correspondence between the experimental barrier gate RB controlling $\Gamma_\mathrm{S}$, and the calculated $\Gamma_\mathrm{S}$ in the phase diagram. The magenta stippled lines in the bottom panels serve as guides to the eye to indicate where the phase transitions occur.

\subsection{Magnetic field characterization}
	\begin{figure}[ht]
		\includegraphics{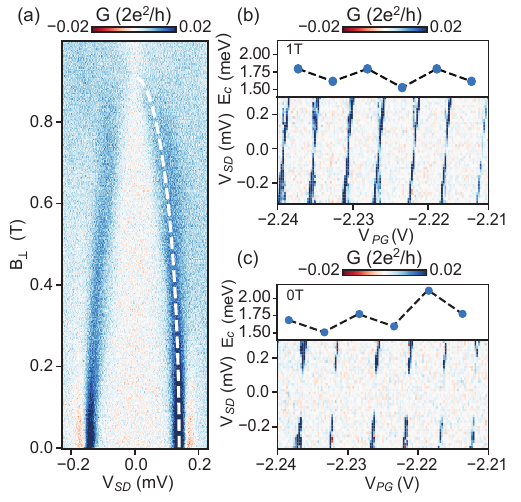}
		\caption{\textbf{(a)]} Bias spectroscopy as a function of out-of-plane magnetic field $B_{\perp}$, {with $V_\mathrm{PG}=-2.225$~V, $V_{HG}=1.000$~V, $V_\mathrm{RB}=-1.450$~V and $V_\mathrm{LB}=-0.825$~V}. We extract a critical field $B_{c,{\perp}}$ = 0.90$~\pm~0.04$~T. \textbf{(b-c)} Bias spectroscopy at low $\Gamma_\mathrm{S}$ {$V_\mathrm{RB}=-1.300$~V, $V_\mathrm{LB}=-0.825$~V, and $V_{HG}=-1.000$~V  } for out-of-plane magnetic field $B_{\perp}= 1$~T, and $B_{\perp}= 0$~T respectively. Top panels show extracted charging energies from Coulomb diamonds below.}
	\label{fig:Figure.3}
	\end{figure}

\begin{figure*}
    \includegraphics{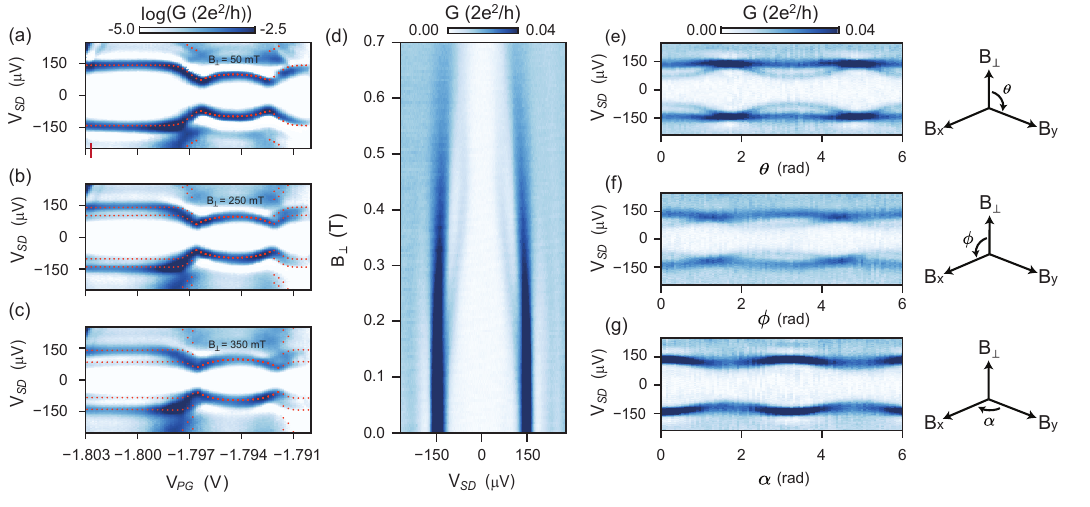}
  \caption{\textbf{(a-c)} Bias spectroscopy displayed in logarithmic scale at magnetic field strengths of {50~mT}, 250~mT and 350~mT respectively, {with the barrier gates set to $V_\mathrm{RB}=-1.4100$~V, $V_\mathrm{LB}=-0.8070$~V and $V_{HG}=-0.9575$~V.} Sub-gap energy splitting is visible in b) and c). The superimposed orange markers in Figures a-c serve as a qualitative comparison between the measured transport data and calculations using a zero-bandwidth model, inputting a {$\Delta=71\,\,\upmu $eV}, hybridization energy $\Gamma_\text{S}$=110~$\upmu $eV, charging energy $U$ = $1.6$ meV, as well as a Zeeman energy of {$E_z^{SC}=4~\upmu$eV} on the SC and {$E_z^{QD}=4~\upmu$eV} on the QD, $E_z^{SC}=44\,\, \upmu $eV on the SC and $E_z^{QD}=44 \,\,\upmu $eV on the QD, and $E_z^{SC}=56 \,\,\upmu $eV on the SC and $E_z^{QD}=56 \,\,\upmu $eV on the QD, respectively. (d) Magnetic field sweep {at $V_\mathrm{RB}=-1.4100$~V, $V_\mathrm{LB}=-0.8070$~V, $V_\mathrm{HG}=-0.9575$~V, and $V_\mathrm{PG}=-1.8027$~V tuned within the singlet ground state, as indicated by the red notch in a)}. We extract a g-factor for the out-of-plane magnetic field splitting $g_{\perp}$ = $1.5\pm0.2$. (e-g) Bias spectroscopy of rotating magnetic field at total magnetic field strength $|B|$ = 420~mT {and $V_\mathrm{PG}=-1.8028$~V, also tuned to the singlet groundstate}. Strong $g$-tensor anisotropy is observed between out-of-plane and in-plane magnetic field orientations. }
    \label{fig:Figure.4}
\end{figure*}

We now turn to the magnetic field dependence of the superconducting parent gap. Figure \ref{fig:Figure.3}a shows bias spectroscopy of the quantum dot in the few hole and low $\Gamma_\mathrm{S}$ regime, as a function of out-of-plane magnetic field strength $B_\perp$. We fit the closing of the superconducting gap according to $\Delta(B)$ = 2$\Delta_0 \sqrt{1-(B/B_c)^{2}}$~\cite{tinkham}, where $\Delta_0$ is the superconducting gap at zero magnetic field, and $B_c$ is the critical magnetic field. We extract a critical field of {$B_{c,\perp}$=$0.90~\pm ~0.04$~T}. This greatly exceeds the critical field measured in prior studies, where the critical out-of-plane magnetic field was measured to be approximately 50~mT~\cite{Tosato2023}. This disparity could be due to the smaller size of the superconducting junctions and leads~\cite{Thin_films} measured in the present experiment compared to those in ref.~\cite{Tosato2023}, which may alleviate vortex formation {(see Supplementary section IV.)}. We also characterize the critical magnetic field strengths for the two in-plane axes in Supplementary Section III. Figures \ref{fig:Figure.3}b and \ref{fig:Figure.3}c show bias spectroscopy of the quantum dot in the low coupling limit, taken at 1~T and 0~T respectively. At 0~T the superconducting gap is present within the Coulomb diamonds (Figure \ref{fig:Figure.3}b), while normal Coulomb diamonds are recovered at 1~T due to the breaking down of superconductivity (Figure \ref{fig:Figure.3}c). In both cases we find an even-odd oscillation in the filling structure at both low and high field strengths as seen in the top panels of Figures \ref{fig:Figure.3}b and \ref{fig:Figure.3}c, depicting the addition energy of each Coulomb diamond below, consistent with that observed previously in germanium~\cite{vanRiggelen2021} and InAs~\cite{Grove-Rasmussen2009} QDs. The charging energy of the quantum dot varies between even and odd periodicity, indicating that the quantum dot is in the low hole occupancy. The charging energy is typically between 1 and 1.8~meV, more than ten times the superconducting gap, supporting our interpretation that we are in the YSR regime.\\

Finally we study the transport spectrum under the influence of a magnetic field in the same electrostatic regime as Figure \ref{fig:Figure.2}b-c. Figures \ref{fig:Figure.4}a-c show bias spectroscopy measurements as a function of $V_\mathrm{PG}$ portrayed in logarithmic scale for enhanced visibility, taken at a perpendicular magnetic field, at strengths of (a) {50}~mT, (b) 250~mT, and (c) 350~mT (see Supplementary Figures S7a-b for data in non-logarithmic scale). At {50~mT} a $\ket{S}$-$\ket{D}$-$\ket{S}$ transition spectrum is observed, as seen in Figure 2b-c.  A qualititive ZBW model of the system is plotted on top of the data as described previously (see Supplementary section V for details of the model), using a SC pairing energy of {$\Delta_0=71~\upmu\mathrm{eV}$}, a hybridization energy of $\Gamma_\text{S}=110~\upmu\mathrm{eV}$, and a QD charging energy of $U$ = 1.6~meV. As we increase the perpendicular field an energy splitting of the subgap states is observed in the singlet ground-state sectors \ref{fig:Figure.4}b-c. Interestingly, the energy splitting seen in the even parity groundstates have a flat energy dispersion indicating either a spin splitting of the parent gap itself, or of a strongly coupled sub-gap state, which we attribute to spinful excited quasiparticles. To qualitatively model the data, we introduce a Zeeman splitting term into the ZBW model for both the SC and QD. We find that setting the g-factor of the QD and superconducting orbitals to be equal in magnitude is sufficient to phenomenologically model the data potentially due to a g-factor renormalization as a result of hybridization~\cite{deMoor2018,Antipov2018}. Additional data at lower magnetic field strength and lower coupling, supporting our hypothesis of g-factor renormalization, is reported in the Supplementary section II, as well as results from our ZBW model with varying system parameters in Supplementary section V.

In Figure \ref{fig:Figure.4}d, we set the plunger gate voltage {to $V_\mathrm{PG}=-1.8027$ such that the quantum dot is in a singlet ground state as indicated by the red notch} in Figure \ref{fig:Figure.4}a and perform bias spectroscopy as a function of $B_{\perp}$ from 0~T to 0.7~T. We observe the magnetic field splitting of the superconducting coherence peaks, and extract an out-of-plane g-factor, $g_{\perp}$ = $1.5~\pm~0.2$. This value is several times larger than the g-factor measured for magnetic fields close to in-plane in planar germanium quantum wells \cite{Hendrickx2020a,Hendrickx2020b}, and several times smaller than the value reported in ref.~\cite{Hendrickx2023} for out of plane g-factors. Furthermore, in a regime of lower coupling for the YSR states, we have measured an out-of-plane g-factor of $g_\perp=4.5~\pm~0.6$ (see Supplementary Information, Figure S1) and a $g_\perp=5.3~\pm~0.8$ in out-of-gap Coulomb diamond spectroscopy (see Supplementary Information, Figure S2). These observations support the hypothesis that the g-factor $g_\perp$ either describes the quasiparticle coherence peaks, or is due to renormalization of the Ge/SiGe hole $g$-factor as a result of hybridization with the superconductor~\cite{deMoor2018,Antipov2018}. \\

Finally, we investigate anisotropy of the g-factor by performing bias spectroscopy as a function of magnetic field orientation. Figures \ref{fig:Figure.4}e-g show bias spectroscopy at a magnetic field strength of {$|B|=420$~mT}. The splitting is then investigated as a function of rotation angles. Here, the rotation angles $\theta$, $\phi$ and $\alpha$ are defined as shown in Figure \ref{fig:Figure.4}e-g. A large anisotropy is measured; while g-tensor anisotropy is ubiquitous for heavy holes in strained planar Ge/SiGe wells, it is seldom observed for a superconducting gap edge. On the other hand, anisotropic g-tensors have been observed in heavy fermion bulk superconductors, which could additionally explain the non-dispersive splitting in Fig.~4b-c~\cite{chandra2013hastatic}. We stress that our use of the annealed poly-crystalline superconductor PtGeSi is a recent material development in itself, and the physics of superconductivity in these nanoscale thin films is not fully understood. However, this anisotropy could also be explained by a sub-gap state in the QD that is strongly coupled to the superconductor.   

%:Discussion		
\section{Conclusion}
We have demonstrated a quantum dot in Ge/SiGe proximitized by a superconducting lead and exhibiting clear YSR states. We find that the coupling between quantum dot and superconducting lead is highly tunable, as evidenced by the tunnel barrier controllability of the singlet or doublet nature of the YSR ground state at half-filling. Additionally, we have characterized the critical magnetic field strength, finding an out-of-plane critical field of {$\sim0.90$~T}. Finally, we observe Zeeman splitting of sub-gap states, which we explain using a modified zero-bandwidth Anderson impurity model. The ability to tune and strongly couple a superconductor to a quantum dot, in combination with a critical magnetic field, demonstrates the feasibility of our platform for hybrid germanium quantum information processing, including Andreev spin qubits and topological quantum computing, as well as the exploration of fundamental physics with superconductor-semiconductor devices. While the in-plane g-tensor component of holes in germanium planar wells is lower than in established group III-V platforms, it could be enhanced by confinement-induced g-factor engineering~\cite{Bosco2024}, and the out-of-plane g-factor could provide an alternative route to engineering topologically protected qubits using QD-SC chains~\cite{Laubscher2024}. Further work on RF-reflectometry-based measurements (as described in the Supplementary Information section I and II) could help achieve charge sensing and parity readout. {Additionally, studying the dependence of the parent gap and induced gap as a function of temperature and coupling strength would elucidate the nature of the superconducting proximitization of PtSiGe on Ge/SiGe quantum dots~\cite{Morten}. Finally, our work motivates the exploration of alternative materials beyond platinum that also form superconducting germanide phases, such as iridium, rhodium~\cite{RevModPhys.35.1}, and niobium~\cite{langa2024}, with the goal of identifying a germanosilicide capable of hosting large superconducting gap energies while being compatible with the
fabrication process of Ge/SiGe heterostructures~\cite{Lawrie2020}}. Our demonstration in a group IV material, amenable to isotopic purification, constitutes a crucial building block for superconducting-semiconducting hybrid technologies and opens up previously inaccessible experimental directions.
\newline
\newline
$^{\dagger}$ These authors contributed equally.
%:Author contributions	

\section{Author Information}
L.E.A.S and G.S. grew and supplied the Ge/SiGe heterostructures and developed processes for the PtGeSi contacts. L.L. and W.I.L.L. designed the devices. L.L. fabricated the devices. L.L. and W.I.L.L. performed the experiment in the dilution refrigerator along with D.v.D. and {Y.S.}, and analysed the data. L.L. performed numerical simulations within the ZBW model with help from D.v.D and W.I.L.L. L.L and W.I.L.L. wrote the manuscript with input from D.v.D., M.V., G.S., F.K. and A.C. A.C. and F.K. supervised the project.

%:Acknowledgments		
\section{Acknowledgments}
We thank J.~Paaske, V.~Baran and G. Mazur for valuable discussions.
This project has received funding from the European Research Council (ERC) as part of the project NONLOCAL under grant agreement No 856526, and through the IGNITE project under grant agreement No. 101069515 of the Horizon Europe Framework Programme. AC acknowledges support from the Inge Lehmann Programme of the Independent Research Fund Denmark. We acknowledge funding by the Casimir PhD Travel Grant.

\section{Data Availability}
Raw data and analysis scripts for all data included in this work are available at the Zenodo data repository at \url{https://doi.org/10.5281/zenodo.14237057}.


\newpage
\section{Methods}
\subsection{Fabrication}
The device is fabricated on a Ge/SiGe heterostructure, which is grown on n-type Si(001) substrate using a reduced pressure chemical vapor deposition reactor. The stack comprises of a reverse graded Si$_\mathrm{0.2}$Ge$_\mathrm{0.8}$ virtual substrate, a 16~nm Ge quantum well, a 27~nm Si$_\mathrm{0.2}$Ge$_\mathrm{0.8}$ barrier, and a less than 1~nm Si sacrificial cap~\cite{Sammak2019}. The device is fabricated in three electron-beam lithography defined layers separated by two oxide layers of $\sim$~7~nm Al$_2$O$_3$ {grown with a Savannah Ultratech atomic layer deposition system at {$150^{\circ}$C}, using unheated Trimethyl Aluminium~(TMA) percursor and unheated $H_2O$ thermal co-reactant.}~The first layer forms the PtGeSi contacts {defined by electron beam lithography using Kayaku 495 PMMA A2 polymer resist and AR 600-55 1:3 MIBK:IPA developer. Development is directly followed by a 60~s $100^{\circ}C$ post bake on a hotplate before} a wet etch is performed using buffered HF ($\sim 6\%$) solution to remove the sacrificial cap before a 15~nm Pt layer is deposited via e-gun evaporation at a pressure of $1\cross10^{-7}$~mbar. {Lift-off is done in N-Methyl-2-Pyrrolidone~(NMP), followed by 1,2-Dioxolane, and propan-2-ol~(IPA) cleaning.} The Pt contacts then undergo rapid thermal annealing at $400^{\circ}$C for 15~min in {an} argon atmosphere.
Barrier gates and plunger gates are {defined by electron beam lithography using AR-P 6200.04 polymer resist and N-amyl acetate developer, followed by an oxygen plasma ash. The barrier and plunger gates are deposited using electron-gun} evaporation and consist of a Ti~(5~nm) sticking layer and a Pd layer~(25~nm and 29~nm for the barrier and plunger layers respectively). {Lift-off is done in 1,2-Dioxolane, followed by 1,2-Dioxolane, propan-2-ol~(IPA) and oxygen plasma ash cleaning.}
\subsection{Measurement}
The device is measured inside a sample puck loaded into a Bluefors XLD dilution refrigerator, at a mixing chamber temperature of $\sim$~9~mK. The dilution refrigerator is equipped with a (1-1-6)~T vector magnet, the sample chip being placed such that the 6~T direction is in-plane. We expect the main source of magnetic field misalignment to result from misalignment of the sample board inside the puck (QDevil QBoard sample holder) during sample loading. We estimate this error to be less than $5^{\circ}$.
Differential conductance measurements are taken using a Stanford-Research Systems SR860 lock-in amplifier at a frequency of 119~Hz, and amplitude of 2.5~$\upmu V$ {for all figures, except Figure 2a~($V_{AC}=10~\upmu$V), and Figure 4~($V_{AC}=3.2~\upmu$V)}. A line resistance of {$\sim$5.8~k$\Omega$} is subtracted in all measurements. 
Bias spectroscopy measurements contain a small thermally induced bias offsets {on the order of 20~$\upmu$eV, which have been calibrated for and subtracted from all relevant datasets}. DC gate voltages are applied via a QDevil QDAC-II voltage source. {Additonal information on the DC setup and components used can we found in Supplementary Information section I.}
In addition to lock-in measurements, radiofrequency reflectometry was performed as described in the Supplementary Information section I, using a FPGA with a built in microwave signal generator as well as signal demodulator (Quantum Machines OPX+). A tank circuit with a resonance frequency of 192.3~MHz was used, connected to the drain contact. The incident rf-signal is attenuated by 40~dB at various plates of the cryostat, and undergoes an additional attenuation of 20 dB from a directional coupler at the MXC plate (See Supplementary Section Figure {I}). The reflected signal undergoes 40~dB of amplification at the 4 K stage via a HEMT amplifier, and passes through a DC block and bias-tee at the input of the demodulation circuit. A 50 $\Omega$ terminator is connected to the DC side of the bias-tee. 

\end{document}

\title{Supplementary Information:\\ A quantum dot in germanium proximitized by a superconductor}

\author{Lazar Lakic$^{\dagger}$}
\affiliation{Center for Quantum Devices, Niels Bohr Institute, University of Copenhagen, 2100 Copenhagen, Denmark}

\author{William Iain L.~Lawrie$^{\dagger}$}
\affiliation{Center for Quantum Devices, Niels Bohr Institute, University of Copenhagen, 2100 Copenhagen, Denmark}

\author{David van Driel}
\affiliation{
QuTech and Kavli Institute of Nanoscience, Delft University of Technology, Delft, The Netherlands}

\author{Lucas E.~A.~Stehouwer}
\affiliation{
QuTech and Kavli Institute of Nanoscience, Delft University of Technology, Delft, The Netherlands}

\author{Yao Su}
\affiliation{Center for Quantum Devices, Niels Bohr Institute, University of Copenhagen, 2100 Copenhagen, Denmark}

\author{Menno Veldhorst}
\affiliation{
QuTech and Kavli Institute of Nanoscience, Delft University of Technology, Delft, The Netherlands}

\author{Giordano Scappucci}
\affiliation{
QuTech and Kavli Institute of Nanoscience, Delft University of Technology, Delft, The Netherlands}

\author{Ferdinand Kuemmeth}
\affiliation{Center for Quantum Devices, Niels Bohr Institute, University of Copenhagen, 2100 Copenhagen, Denmark}

\author{Anasua Chatterjee}
\email{anasua.chatterjee@tudelft.nl}
\affiliation{Center for Quantum Devices, Niels Bohr Institute, University of Copenhagen, 2100 Copenhagen, Denmark}
\affiliation{
QuTech and Kavli Institute of Nanoscience, Delft University of Technology, Delft, The Netherlands}

\date{\today}
\maketitle

\clearpage
\newpage
\onecolumngrid

\appendix
\renewcommand{\thesubsection}{\Roman{subsection}} 
\setcounter{figure}{0}
\renewcommand\thefigure{S\arabic{figure}}

In the Supplementary Information we present additional data as well as a minimal model for our hybrid system. {In section I, we present details on the experimental setup ,describing both DC and RF measurements. In section II,} we present data in a low regime of coupling between the superconductor (SC) and quantum dot (QD) that showcases the tunability of our system and complements the results presented in the main article. We also present similar results measured using radiofrequency (RF) reflectometry, a technique ubiquitous for spin-based qubits, crucial to develop readout for hybrid superconductor-semiconductor devices incorporating quantum dots, such as Kitaev chains. Section III describes critical field measurements for different orientations of the applied magnetic field. {In Section IV we present data on PtSiGe Hall bars and wires providing an insight into the SC properties of PtSiGe films}. Finally, Section V
presents a qualitative model of the observed spectrum of our proximitized quantum dot.

\section{Section I. Details of Experimental Setup}
In this  section we describe the experimental setup used for both the low-frequency (lock-in) measurements in the main text, and the RF-reflectometry setup used for the additional data presented in the supplement (Section II). 

\begin{figure*}[ht]
		\includegraphics{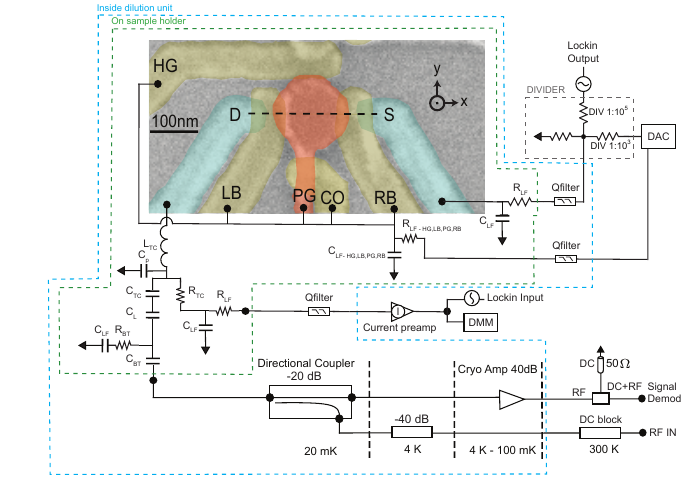}
		\caption{False coloured SEM of a nominally identical device, including the experimental setup used for DC, differential conductance and RF measurements.}
		\label{fig:Supp_fig1}  
   \end{figure*}

All measurements in Figures 1-4 in the main text are performed using a two-terminal measurement setup, measuring a DC current and/or the differential conductance $G$. For two-terminal measurements in which a DC current is measured, we apply a voltage on the source (S) via a QDevil DAC with a 16-bit resolution, and a $\pm$10~V output voltage. This voltage passes through a 1:1000 voltage divider on the DAC output. The  voltage is applied to a DC-loom via a QFilter-II RC low pass filter  ($f_{co}$=65~kHz) and LC~low-pass filter ($f_{co}$=225~MHz), followed by a QDevil sampleboard incorporating further RC low-pass filtering ($f_{co}= 130$~kHz, $R_\mathrm{LF}=1.2~$k$\Omega$, C$_\mathrm{LF}=1$~nF). The wirebond connected to the drain contact connects to the PCB and branches into an RF-reflectometry setup, which is described below, and a DC readout line. The DC readout line on the drain~(D) then passes through the same filtering setup as described above. Outside the cryostat, the voltage is converted to a current via a Basel SP983c current pre-amplifier set to a gain of 1$\times10^{-7}$~V/A before being digitized by a Keysight 34461A digital multimeter. The total line resistance of the sample board, QFilter-II, and DC looms amount to a line resistance of $R_{line}\sim5.80\pm0.03\,~{k\Omega}$. 

The differential conductance measurements are performed by applying an AC excitation via a Stanford Research Systems SR860 lock-in amplifier. This excitation is combined with the DC signal via a 1:10000 voltage divider which otherwise undergoes the same filtering and amplification as the DC current. We furthermore subtract the line resistance of $\sim$5.8k$\Omega$ from each point on the $V_\mathrm{SD}$ axis, $V_\mathrm{SD} = V_\mathrm{DAC} - (I_\mathrm{DC}\cdot R_\mathrm{line}$), where $V_\mathrm{DAC}$ is the voltage being output from the DAC, $I_\mathrm{DC}$ is the measured current, and $R_\mathrm{line}$ the line resistance. We optimize and measure the  in-phase component and convert the signal to units of quantum of conductance by $G = {G_0}^{-1}\cdot({\frac{V_\mathrm{AC}}{X\cdot \mathrm{Gain}}-R_\mathrm{Line})^{-1}}$, with $V_\mathrm{AC}=2.5$ $\upmu$V for all figures, except Figure 2a~($V_{AC}=10~\upmu$V), and Figure 4~($V_{AC}=3.2~\upmu$V), $X$ being the measured in-phase signal in units of
volts, the Gain being 1$\times10^{-7}$~V/A, and $G_0=7.748091719\times10^{-5}$~S. 

Furthermore, we incorporate high-frequency measurement via standard RF-reflectometry methods~\cite{Vigneau2022} using a cryogenic tank circuit with a frequency of 192.3~MHz (as seen in the measurements in the following section, in Figure~\ref{fig:Supp_fig3}). A typical RF excitation voltage at the sample is 3~$\upmu$V, which is applied and demodulated  using a Quantum Machines OPX+. The RF-reflectometry setup has a bias-tee with resistance $R_\mathrm{BT}=50\,k \Omega$, capacitance $C_\mathrm{BT}=22\,n\mathrm{F}$, and an inductance $L_\mathrm{TC}=560$~nH. $C_\mathrm{P}$ is a parasitic capacitance that together with $L_\mathrm{TC}$ forms the resonant (tank) circuit, estimated to have a value of about 1.2~pF. The capacitances $C_\mathrm{TC}=100$~pF and $C_\mathrm{L}=100$~pF are coupling capacitances.
\color{black}

\section{Section II. Low $\Gamma_\mathrm{S}$ B-field Studies in Transport and RF-Reflectometry}
We study the magnetic field dependence of the subgap states in a regime of low coupling between SC and QD, complementary to the strong-coupling regime in the main text.

\begin{figure*}[ht]
		\includegraphics{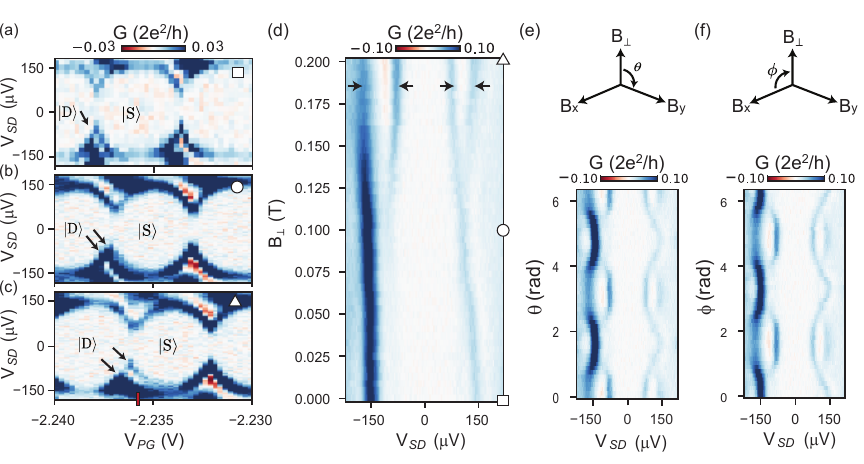}
		\caption{ \textbf{(a-c)} Bias spectroscopy with tunnel barriers set to {$V_\mathrm{RB}=-1.39~V$}, {$V_\mathrm{LB}=-0.8350~V$}, and {$V_\mathrm{HG}=-1~$V} at 0~mT~(square in (a)), 100~mT~(circle in (b)) and  200~mT~(triangle in (c)). Black arrows indicate spin splitting at the transition between a $\ket{D}$ groundstate and a $\ket{S}$ groundstate. \textbf{(d)} Magnetic field sweep at {$V_\mathrm{PG}=-2.2365$~V} ({red} notch in (c)), showing spin-splitting of the odd-occupied state. We extract an out of plane g-factor of $g_{\perp}$ = $4.5\,\pm\,0.6$. \textbf{(e-f)}  Bias spectroscopy as a function of magnetic field angle at total magnetic field strength $|B|$ = 200~mT and {$V_\mathrm{PG}$=-2.2365~V}. Above each panel, we state the rotational angle with respect to the sample direction. A strong $g$-tensor asymmetry is observed for the QD between out-of-plane and in-plane magnetic field orientations. }
		\label{fig:Supp_fig2}  
   \end{figure*}

Setting the tunnel barriers to {$V_\mathrm{RB}=-1.39$~V} and {$V_\mathrm{LB}=-0.8350$~V}, we tune the QD into a many-hole regime using the plunger gate PG. We perform bias spectroscopy at 0~mT, observing filling of the QD with eye-like features characteristic of YSR states as described in the main text, and shown in Figure \ref{fig:Supp_fig2}a. In Figure \ref{fig:Supp_fig2}b the out-of-plane magnetic field is set to 100~mT, revealing lifting of what we believe is subgap spin degeneracy of the $\ket{D}$ groundstate, as indicated by the black arrows. 

In Figure \ref{fig:Supp_fig2}{c} the splitting is increased as shown by the black arrows, when we apply a field of 200~mT. We then configure the PG to {$V_\mathrm{PG}=-2.2365$~V} at the subgap splitting in Figure  \ref{fig:Supp_fig2}c (orange notch), and vary the out-of-plane magnetic field as seen in Figure \ref{fig:Supp_fig2}d. A Zeeman splitting is measured in bias spectroscopy starting from 0~mT. Extracting the g-factor using the standard the Zeeman formula yields a $g_{\perp}$ = $4.5\,\pm\,0.6$, consistent with previous experimental findings of large g-factors in planar Ge/SiGe, and $\sim 3$ times larger than the g-factor we found for the strongly hybridized system described in the main text in this study. As the magnetic field is rotated along the principal axis at a perpendicular field of 200~mT a strong anisotropy is seen, in agreement with previous studies carried out in planar Ge/SiGe.

\begin{figure}
		\includegraphics{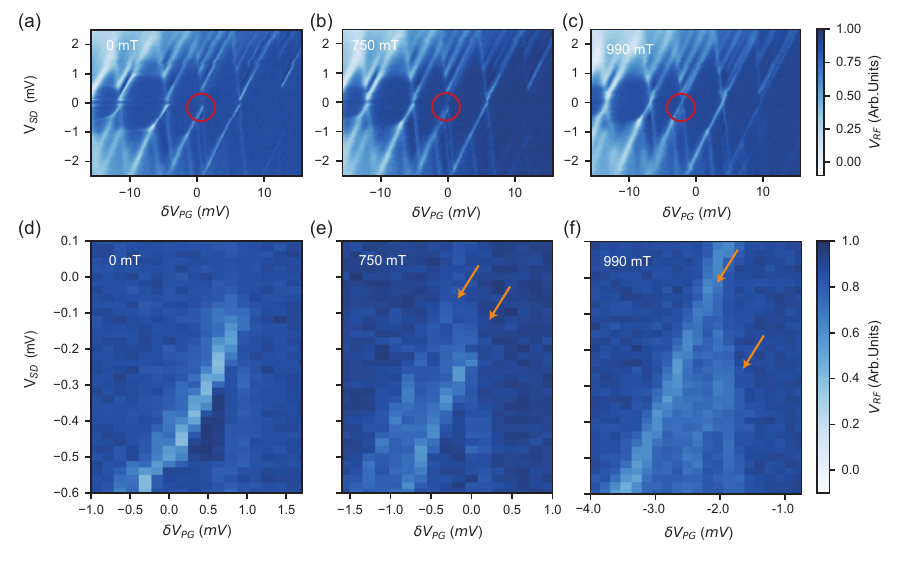}
            \caption{ \textbf{(a-c)} Bias spectroscopy, measured using RF reflectometry, at 0~mT~(a) {with electrostatic gates set to $V_\mathrm{RB}=-1.48~$, $V_\mathrm{LB}=-1.11~V$, and $V_\mathrm{HG}=0~$V}, 750~mT~(b) and 990~mT~(c). (d-f) Zoomed-in scan within the region marked by {red} circles in panels c-e respectively. Orange arrows indicate spin splitting, with a g-factor of $g=5.3 \pm 0.8$.  }
            \label{fig:Supp_fig3}
   \end{figure}

We proceed to tune the QD to a few-hole regime and perform bias spectroscopy using standard RF-reflectometry methods as described above. Note {that the Figures in ~\ref{fig:Supp_fig3} 
show} the normalized demodulated signal $V_\mathrm{RF}$.

Figures \ref{fig:Supp_fig3}c-e depict Coulomb diamonds measured at magnetic field strengths of 0~mT, 750~mT and 990~mT respectively, in the few-hole, lower-coupling regime described in this Supplementary Information section. Suppression of transport exists in an energy window of gap of $\pm2\Delta_0$, as well as Coulomb diamonds with charging energies exceeding 2~mV. The x-axis is labeled with a virtual parameter $\delta V_\mathrm{PG}$, where $\delta V_\mathrm{PG}=0$ corresponds to {$V_\mathrm{PG}=-2$~V}  sweeping 15~mV above and below.
Figures \ref{fig:Supp_fig3}c-e depict bias spectroscopy from which we extract $g_\perp$ outside the superconducting gap. Figures \ref{fig:Supp_fig3}f-h correspond to zoomed-in plots of figures \ref{fig:Supp_fig3}c-e, revealing a magnetic field dependent splitting. The extracted g-factor is $g=5.3 \pm 0.8$, in agreement within error with the subgap spin splitting that we measured using low-frequency transport techniques, in Figure \ref{fig:Supp_fig2}.

\clearpage
\section{Section III. Measurements of Critical Field}

\begin{figure*}[ht]
		\includegraphics{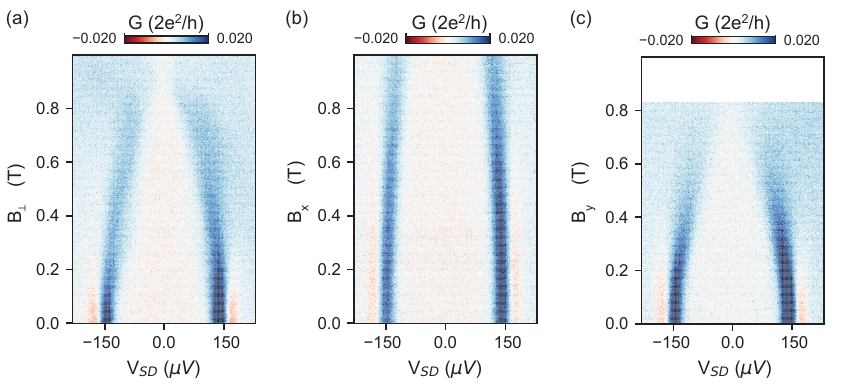}
		\caption{{Magnetic fields sweeps at same electrostatic gate settings as in main Figure 3a.} \textbf{(a)} Magnetic field sweep from 0 to 1~T out of plane with respect to the device. \textbf{(b)} Magnetic field sweep from 0 to 1~T for in-plane field, parallel to the direction of transport \textbf{(c)} Magnetic field sweep from 0 to 0.82~T for in-plane field, perpendicular to the direction of transport.}
            \label{fig:Supp_fig4}
\end{figure*}

We perform bias spectroscopy as a function of magnetic field in Figure \ref{fig:Supp_fig4}a-c, with the device configured in the few-hole regime. Figure \ref{fig:Supp_fig4}b shows the critical in-plane field parallel to the direction of transport (see Figure \ref{fig:Supp_fig1}a for a definition of the x direction). This configuration is expected to have a higher critical field than the perpendicular magnetic direction due to a smaller surface area, thus making it harder for the magnetic field flux to pin and destroy superconductivity. Finally, Figure \ref{fig:Supp_fig4}c depicts the critical in-plane field perpendicular to the direction of transport (see Figure \ref{fig:Supp_fig1}a for a definition of the y direction). The critical field in this direction is marginally lower than the out-of-plane magnetic field. 

\clearpage
\section{Section IV. Electrical Characterization of PtSiGe Films}
\begin{figure*}
		\includegraphics{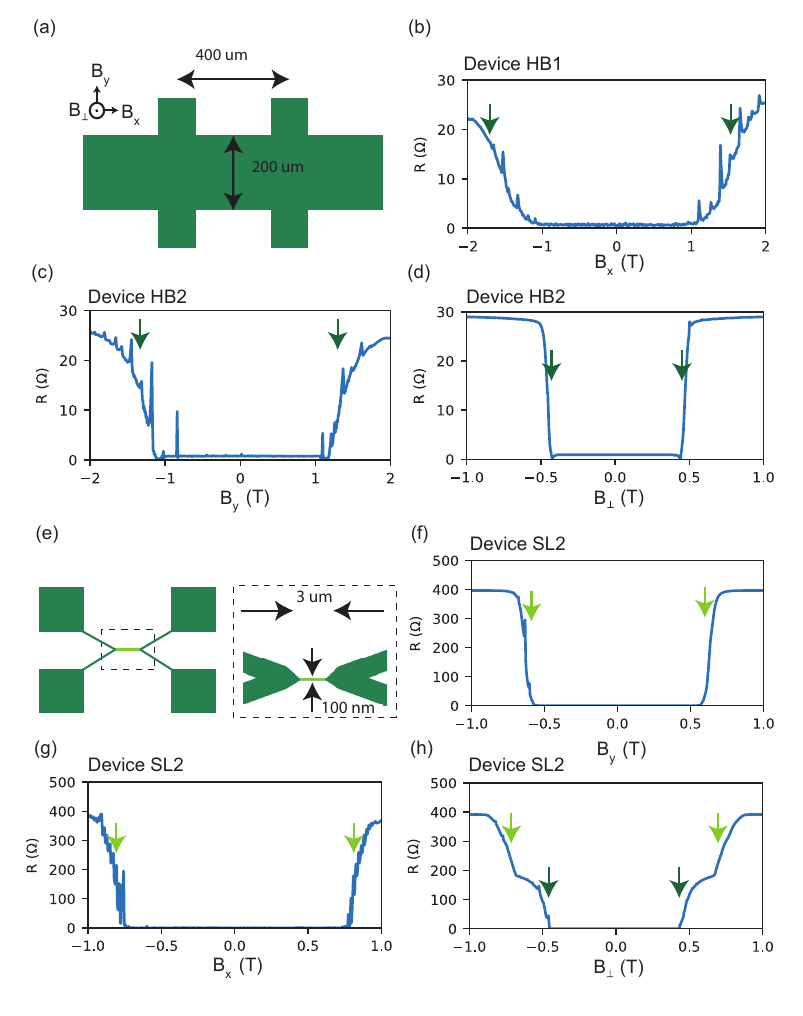}
		\caption{\textbf{(a-d)} Critical magnetic field measurements for PtSiGe Hall bars. (a) Schematic of Hall bar geometry and alignment of magnetic field directions. Here, $B_{x}$ indicates a magnetic field applied in the plane of the heterostructure, aligned with the direction of transport, $B_{y}$ indicates a magnetic field applied in the plane of the heterostructure, perpendicular with the direction of transport, and $B_{\perp}$ indicates a magnetic field applied out-of-plane with respect to the heterostructure. Dark green arrows represent the magnetic field at which a superconducting-normal state transition occurs according to the fit. \textbf{(e-g)} Critical magnetic field measurements for PtSiGe wires. (e) Schematic of wire geometry. The wires are 3 $\upmu$m long, 100 nm wide, with a Pt deposition thickness of 15~nm. In (h), two critical magnetic field onsets occur due to the voltage probes being sensitive to voltage drops in part of the fanout (inset of (e)). The dark green arrows represent the superconducting-normal state transition of the larger `fanout' section, while the light green arrows represent the superconducting-normal state transition of the wires. }
            \label{fig:Supp_fig5}
\end{figure*}

To study superconducting properties, we also perform measurements of the critical magnetic field strengths, for PtSiGe Hall bars and wires fabricated using identical methods and on the same wafer as the device reported in the manuscript. Figure \ref{fig:Supp_fig5}a shows a schematic of the Hall bar dimensions and magnetic field orientation. We perform current bias measurements using a lock-in amplifier at 33 Hz excitation frequency and 0.5 V amplitude, applied through a 1 M$\Omega$ resistor. Figures \ref{fig:Supp_fig5}b-d show measurements of differential resistance as a function of applied magnetic field strength for two separate Hall bars HB1 and HB2. Differential resistance is calculated by dividing the in-phase lock-in signals of the longitudinal voltage drop $V_{xx}$ by the measured current at the drain electrode $I$. We fit a sigmoid function to the data of the form $R(B)=R_N/(1+f(B))$ where $f(B) = \exp [-(B-B_c)/c]$. Here, $R_N$ is the normal state resistance, $B^c$ is the critical magnetic field, and $c$ represents the broadening of the Normal-Superconducting transition. Critical magnetic fields for each Hall bar are reported in Table 1. As expected, the out-of-plane critical field $B^c_{\perp}$ is smaller compared to the in-plane fields for a Hall Bar geometry, as observed in ref~\cite{Tosato2023}.

We also perform measurements of Critical magnetic fields on wires, whose geometries are shown in figure \ref{fig:Supp_fig5}e. Two identical devices S1 and S2 are studied. Critical magnetic fields are reported in Table 1. We note that in figure S5h, there appear to be two transitions as shown using light and dark green. This is likely due to the design of the device, where the voltage probes will detect drops across part of the fanout, which have a much larger surface area than the wires themselves, resulting in a much smaller critical field. This is further evidenced when comparing the critical fields of the Hall bar devices and the wires, whereby the wires typically exhibit much higher out-of-plane critical magnetic fields.

\begin{figure*}
		\includegraphics{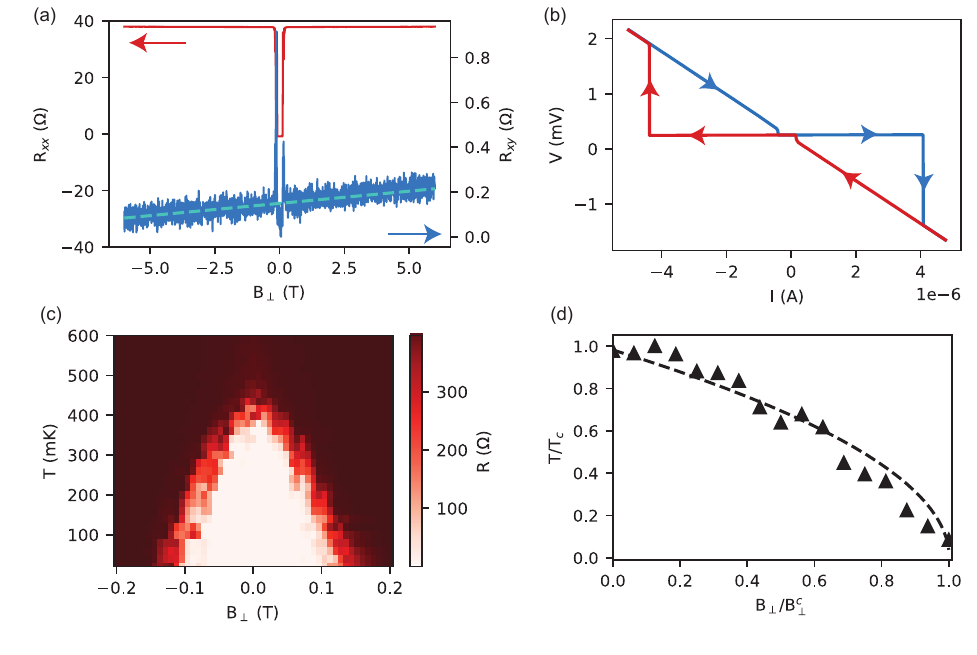}
		\caption{\textbf{(a)} Longitudinal  (red) and hall (blue) resistance of Hall bar HB3 for a symmetric out of plane magnetic field range of -6 T to 6 T. A linear fit of the Hall resistance of the device in the normal state is shown in the cyan dashed line, resulting in an estimated carrier density of 5.2 $\times 10^{16}$ cm$^{-2}$. We estimate the mobility of the carriers in the PtSiGe Hall bar (aspect ratio L/W = 2) to be 6 cm$^{2}$/Vs. \textbf{(b)} Critical current measurements on a 100 nm wide PtSiGe wire at zero magnetic field and at base cryostat temperature. Arrows indicate the direction of the sweep. A critical current density of 2.7 A/m$^2$ is extracted assuming a PtSiGe thickness of 15 nm. \textbf{(c)} Critical out-of-plane magnetic field as a function of temperature recorded at the mixing chamber plate of the cryostat. Instabilities in PID loop impart approximately a $\pm$10 mK error bar on the temperature axis. A critical temperature of $T_c$ = 408 mK and critical out-of-plane magnetic field of 0.116 T are extracted from sigmoid function fits to the data. \textbf{(d)} Fitted critical temperature and magnetic field from c) scaled by their $B^c_\perp$($T$=0) and $T^c$($B_\perp$=0) values respectively. The dashed line is a BCS theory fit to the data. }
            \label{fig:Supp_fig6}
\end{figure*}

We further study the electrical properties of the PtGeSi material, by measuring longitudinal and Hall resistances as a function of out of plane magnetic field, in Hall bar HB3 (Figure \ref{fig:Supp_fig6}a). Sweeping magnetic field strength from -6 T to 6 T, we observe a linear dependence of the Hall resistance $R_{xy}$ in the normal phase of the material. Fitting this, we extract a sheet carrier density of $n_{\text{2D}}$ = 5.2 $\times 10^{16}$ cm$^{-2}$. We use this value combined with the longitudinal (normal) resistance $R_N$ = 37.9 $\Omega$, to extract a carrier mobility of $\mu$ = 6 cm$^{2}$/Vs.

Figure \ref{fig:Supp_fig6}b shows current bias measurements through a wire device SL1, from which we extract a critical current of 4 $\upmu$A. Assuming a cross section of 100 nm x 15 nm for the wire, we calculate a critical current density of $J$ = 2.7 A/m$^2$. Due to the similarity in feature size, we expect this current density to reflect that of the superconducting leads in the device reported in the main manuscript. We note that measurements in the main text are taken at currents restricted to around 1 nA or lower (Main text Figure 1d) significantly lower than the expected critical current of 2.4~$\upmu$A, and are therefore unlikely to exhibit features due to depairing~\cite{Ibabe2023}.

Finally, we study the critical temperature of the PtSiGe material in a wire device (SL2). We use a PID feedback loop on a heater placed at the Mixing chamber temperature stage of the cryostat, to control the temperature of the Mixing chamber plate. This temperature is stable within a range of $\pm 10$ mK of the setpoint. We note that the temperature of the mixing chamber plate is not necessarily reflective of the temperature of the sample, as the sample is loaded inside a puck via a fast sample exchange unit, and is likely to underestimate the actual temperature by some tens of mK. We perform out-of-plane magnetic field sweeps between -0.2 T and 0.2 T, while stepping the mixing chamber temperature setpoint of the heater in steps of 20 mK, from 20 mK to 600 mK. After each temperature step, a settle time of 5 minutes is allowed before beginning the magnetic field sweep. Figure 6c shows the results of this measurement, indicating a clear switching between superconducting and normal phases. We fit the critical magnetic fields and critical temperatures using a sigmoid function, finding $T_c$ = 408 mK, and $B^c_\perp$ = 0.116 T. We plot the scaled phase diagram in figure \ref{fig:Supp_fig6}d, and fit the standard BCS relation $T/T_c$ = 1.74$\sqrt{1-B_\perp/B^c_\perp}$ with good agreement.
\color{black}
\begin{table}[]
    \centering
    \begin{tabular}{c|c|c|c|c}
          & Device Type & $B^*_{x}$ (T) &$B^*_{y}$ (T) & $B^*_{\perp}$ (T)  \\
         \hline
         HB1     &Hallbar     &1.54(1)     &-      & -\\
         HB2     &Hallbar     &$>$1     &1.34(1)      &0.474(1)\\
         HB3     &Hallbar     &$>$1     &$>$1    & 0.123(6)\\
         \hline
         SL1     &  Wire  &0.831(6)       &$>1$      &0.643(1)  \\
         SL2     &  Wire  &0.827(1)       &0.633(1)      &0.695(1)* \\

    \end{tabular}
    \caption{Summary of measured critical magnetic fields for different Hall bar and wire devices under test. *We note that for wire SL2, two critical magnetic fields emerge, which we attribute to different geometries of the wire and the fanout. The number reported is the higher of the two critical fields. For completeness, the lower critical field occurs at $B_\perp$ = $0.500~\pm~0.001$~T.}
    \label{tab:SLHBB}
\end{table}
\clearpage
\newpage
   \section{Section V. Modeling of Subgap States}

The device in the main text consists of a QD weakly coupled to one superconducting lead, and strongly coupled to another. We therefore write the Hamiltonian of the total system as:
\begin{equation}
    H=H_{QD}+H_T+H_{SC}.
\end{equation}

Where $H_{QD}$ is defined as

\begin{equation}
    H_{QD} = \sum_{\sigma}{\epsilon}{d_\sigma^{\dagger} d_\sigma}+U{d_\downarrow^{\dagger}d_\downarrow d_\uparrow^{\dagger}d_\uparrow}+\frac{E_z}{2}(d_\uparrow^{\dagger}d_\uparrow-d_\downarrow^{\dagger}d_\downarrow )
\end{equation} 
Here, $\epsilon$ is the electrochemical potential, $U$ is the charging energy and $E_Z$ is the Zeeman energy. As one superconducting lead is weakly coupled, we consider only one bulk superconductor for $H_{SC}$, and ignore possible phase differences between their order parameters. In the superconducting Anderson impurity model, $H_{SC}$ consists of a sum over multiple orbitals in the superconductor. Following Refs.~\cite{meng2009self,bauer2007spectral}, we replace this sum by a single orbital and write:
\begin{equation}
H_{SC}=\xi\sum_{\sigma}c_{\sigma}^{\dagger} c_{\sigma} - \Delta(c_{\uparrow}^{\dagger} c_{\downarrow}^{\dagger}+h.c.)+\frac{E^{SC}_z}{2}(c_\uparrow^{\dagger}c_\uparrow-c_\downarrow^{\dagger}c_\downarrow )
\end{equation}
Where $\xi$ is the single-electron energy, $\Delta$ is the parent gap and $E^{SC}_z$ is the Zeeman energy. We set $\xi=0$, as the superconductor is grounded throughout the experiment. The Zeeman splitting of the superconductor is commonly excluded, as metallic superconductors usually have significantly lower g factors than the semiconductors they are coupled to. We however choose not to exclude it, in view of the low in-plane g-factor of Ge/SiGe heterostructure QDs, as well as the little-known magnetic field behavior of the novel PtGeSi superconductor. Finally, we can write the coupling of the superconductor and QD as:
\begin{equation}
H_T=\Gamma_S\sum_{\sigma}(c_\sigma^{\dagger} d_\sigma + h.c.)
\end{equation}
where $\Gamma_S$ is the spin-conserving hybridization energy. The full basis consists of 16 states, which we write as the tensor product of the number states of the superconductor and QD $\ket{SC,QD}$: 
\begin{equation}
\{\ket{0,0},\ket{\uparrow,0},\ket{\downarrow,0},\ket{2,0},
\ket{0,\uparrow},\ket{\uparrow,\uparrow},\ket{\downarrow,\uparrow},\ket{2,\uparrow},
\ket{2,\downarrow},\ket{\uparrow,\downarrow},\ket{\downarrow,\downarrow},\ket{2,\downarrow},\ket{0,2},\ket{\uparrow,2},\ket{\downarrow,2},\ket{2,2}\}
\end{equation}
To model the spectrum, we consider excitations between the even and odd parity states. As the tunnel probe is a superconductor, we offset all excitation energies by $\Delta$.

In Main Text Fig.~4b,c a Zeeman splitting of the superconductor was required to match the model to the experiment. For the model overlays seen in main text Figure 4a-c, the parameters used were a superconducting gap of $\Delta=71~\upmu$eV, a charging energy of $U=$1.6 meV, a hybridization energy of $\Gamma_S$=~110~$\upmu $eV along with an energy splitting on both SC and QD. We phenomenologically include the magnetic field dependence of the parent gap as: $\Delta(B)$ = 2$\Delta_0 \sqrt{1-(B/B_c)^{2}}$, where the factor of $2$ is due to both leads having a superconducting density of states, resulting in quasiparticle transport between the bulk superconductors at  $E=2\Delta$ and $B_c=0.9$ T as the critical field. This is required for agreement between the experimental data and the model, as states are observed at $E_\mathrm{state}+\Delta$ due to the superconducting tunnel probe.

\begin{figure}
    \includegraphics{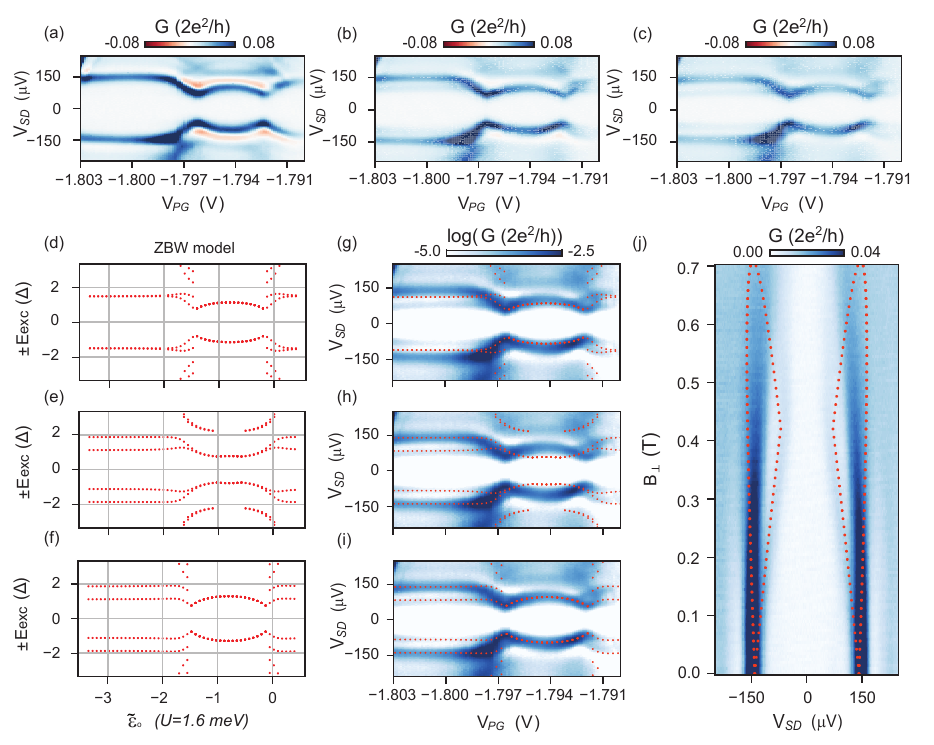}
	\caption{ \textbf{(a-c)} Bias spectroscopy data from main text Figures 4(a-c) without the logarithmic scale. \textbf{(d)} Excitation energy in units of $\Delta=71\upmu $eV as a function of chemical potential in units of charging energy $U=1.6$~meV, with a Zeeman energy only on the QD of $E_z$=56~$\upmu $eV. \textbf{(e)} Excitation energy in units of $\Delta=71\upmu $eV as a function of chemical potential in units of charging energy $U=1.6$~meV, with a Zeeman energy of $E_z$=56~$\upmu $eV on SC only. \textbf{(f)} Excitation energy in units of $\Delta=71~\upmu$eV as a function of chemical potential in units of charging energy $U=1.6$~meV, Zeeman energy on both SC and QD of $E_z$=$56~\upmu $eV. \textbf{(g-i)} Bias spectroscopy data from main Figure 4c, overlaid with the computed model ({red}) from supplement Figure 4d-f respectively. \textbf{(j)} Bias spectroscopy data from main Figure 4d overlaid with the computed ZBW model (orange) correcting for the gap closing as the magnetic field increases using a g-factor of 1.5 on both SC and QD.
            \label{fig:Supp_fig7}}
\end{figure}

Using the described ZBW model we can compute the energy spectrum of the QD hybridizing with the SC.
In \ref{fig:Supp_fig7}d and g, we 
adding a Zeeman energy of $E_z$ = 56~µeV to the QD energy and nothing on the superconductor, which results in a spin-splitting at the charge degeneracy points $\epsilon(\Delta)=0$ and $\epsilon(\Delta)=-1.6$, but no splitting in the even parity regions ($\ket{S}$), contradicting the data in Figure \ref{fig:Supp_fig7}g. In supplement Figure \ref{fig:Supp_fig7}e and h, we add a Zeeman energy of $E_z$=56 $\upmu $eV only to the SC, resulting in a splitting in the even parity states ($\ket{S}$) resembling the energy separation seen in main text Figure 4b-c, however we see no spin-splitting at the charge degeneracy points $\epsilon(\Delta)=0$ and $\epsilon(\Delta)=-1.6$; furthermore, the characteristic eye-like curvature of the $\ket{D}$ state is significantly reduced, and does not match the data in Figure \ref{fig:Supp_fig7}h. In Figure \ref{fig:Supp_fig7}f and i, a Zeeman energy $E_z$=56~$\upmu $eV was added to both the QD and the SC, resulting in qualitative agreement to the data in Figure \ref{fig:Supp_fig7}i. We stress that our model is quite simple and is not meant to fit the data quantitatively, and as such the magnitude and sign of the respective Zeeman terms have not been varied extensively. For example, future theoretical work could include the spin-orbit coupling, as well as a more detailed description of the superconductor beyond zero band width.

Finally, in Fig \ref{fig:Supp_fig7}j we show the bias spectroscopy data from main text Figure 4d, overlaid with the computed ZBW spectrum. There is a good agreement between the model and the data until about 400~mT, when the splitting becomes larger than the SC pairing energy. We attribute this discrepancy between model and data to the ZBW model, which cannot model the Zeeman effect on a bulk superconducting density of states.

\newpage

\begin{figure}
		\includegraphics{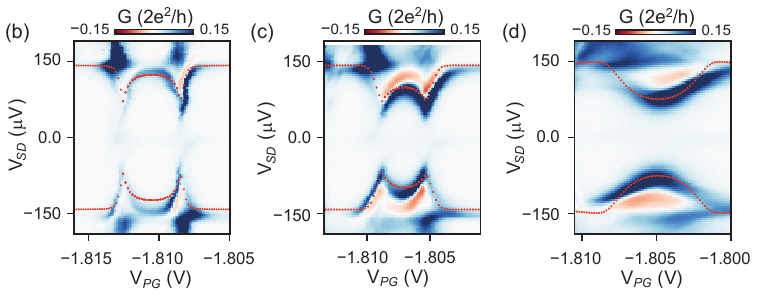}
		\caption{ \textbf{(a)} Bias spectroscopy data from main text Figure 2b overlaid with the computed ZBW model, setting the gap energy to {$\Delta=71\, \,\upmu $eV}, a charging energy of $U=1.6$~meV and a $\Gamma_S=70\,\,\upmu $eV. \textbf{(b)} Bias spectroscopy data from main text Figure 2c overlaid with the computed ZBW model, setting the gap energy to $\Delta=71 \,\,\upmu $eV, a charging energy of $U=1.6$ meV and a $\Gamma_S=110 \,\,\upmu $eV. \textbf{(c)} Bias spectroscopy data from main text Figure 2d overlaid with the computed ZBW model, setting the gap energy to {$\Delta=71\,\,\upmu $eV}, a charging energy of $U=1.6~$ meV and a $\Gamma_S=150 \,\,\upmu $eV.
            \label{fig:Supp_fig8}}
\end{figure}

In order to quantify the hybridization energy of SC and QD $\Gamma_S$, we model the subgap states in main text Figure 2b-d in the ZBW model, using the superconducting gap and charging energy extracted in the main text. We manually tune the parameter $\Gamma_S$ until a qualitative match between data and model is found. For Figures \ref{fig:Supp_fig8}a-c we set the SC gap to {$\Delta=71~\upmu $eV} and the charging energy to $U=1.6~$ meV, for which we extract an estimate of $\Gamma_S=70 \,\,\upmu $eV {for Figures \ref{fig:Supp_fig8}a, $\Gamma_S=110 \,\,\upmu $eV for Figure \ref{fig:Supp_fig8}b, and $\Gamma_S =150 \,\,\upmu $eV for the fully coupled groundstate in Figure \ref{fig:Supp_fig8}c.}

\newpage
$^{\dagger}$ These authors contributed equally.